\documentclass[12pt]{article}
\usepackage{amssymb}

\author{I. G. Korepanov\\
\normalsize Southern Ural State University\\[-0.5ex]
\normalsize 76 Lenin avenue\\[-0.5ex]
\normalsize 454080 Chelyabinsk, Russia\\[-0.5ex]
\normalsize E-mail: kig@susu.ac.ru}
\date{}
\title{Exact solution for a matrix dynamical system with usual and
Hadamard inverses}

\def\be{\begin{equation}}
\def\ee{\end{equation}}

\def\dh{\mathop{\rm dh}\nolimits}

\begin{document}
\maketitle

\begin{abstract}
Let $\mathcal A$ be an $n\times n$ matrix with entries~$a_{ij}$ in the field~$\mathbb
C$. Consider the following two involutive operations on such matrices:
the matrix inversion $I\colon\;\mathcal A \mapsto \mathcal A^{-1}$ and the 
element-by-element (or Hadamard) inversion $J\colon\;a_{ij} \mapsto a_{ij}^{-1}$. 
We study the algebraic dynamical system generated by iterations of the product~$J\circ
I$. In the case $n=3$, we give the full explicit solution for this system
in terms of the initial matrix~$\mathcal A$. In the case $n=4$, we provide an explicit 
ansatz in terms of theta-functions which is full in the sense that it works
for a Zariski open set of initial matrices. This ansatz also generalizes
for higher~$n$ where it gives partial solutions.
\end{abstract}


\section{Introduction}
\label{sec intro}

A composition of two noncommuting involutions acting on square matrices
generates sometimes an interesting dynamical system (more specifically:
an algebraic dynamical system with discrete time). A good example can be
found in paper~\cite{diss} (see also references there), where block matrices
$\pmatrix{A&B\cr C&D}$ were considered, $A$, $B$, $C$ and $D$ being themselves
matrices~$n\times n$. The first involution consisted in taking the usual
matrix inverse (of the whole block matrix), and the second one was the
following block transposing: $\pmatrix{A&B\cr C&D} \mapsto \pmatrix{A&C\cr B&D}$.
Such a system (at least when considered to within some natural gauge freedom) 
was shown in~\cite{diss} to be a typical solitonic system solved by usual 
algebraic-geometrical methods.

There is a natural desire to find solutions to more general (than just
solitonic) dynamical systems. Note that the dynamical system generated
by the following transformation: $z\mapsto z^2$, where $z$ belongs to the
unit circle in the complex plane, is obviously solvable but exhibits (as
much obviously) a chaotic, not solitonic, behavior. 

The aim of this paper is to present solutions to the following algebraic 
dynamical system (see \cite{AMV} about its origin and some results for its
particular cases). Let $\mathcal A$ be a $n\times n$ matrix whose entries~$a_{ij}$
belong to the complex field~$\mathbb C$. We consider two involutive operations on 
such matrices: the matrix inversion $I\colon\;\mathcal A \mapsto \mathcal A^{-1}$ 
and the  element-by-element (or Hadamard) inversion 
$J\colon\;a_{ij} \mapsto a_{ij}^{-1}$. Our dynamical system is generated by 
iterations of the product~$J\circ I$. 

The case $n=2$ is trivial but serves us as a useful warm-up exercise.
For the case $n=3$, we give the full explicit solution for this system
in terms of the initial matrix~$\mathcal A$. For the case $n=4$, we provide 
an explicit ansatz in terms of one-dimensional theta-functions. This ansatz
provides a full solution in the sense that it encompasses a Zariski open set 
of matrices. For the case $n\ge 5$, the same ansatz also works but encompasses
only a subvariety of matrices of nonzero codimension.

\section{$2\times 2$ matrices}
\label{sec 2*2}

Despite the triviality of this system, its solution supplies us with some
tool that will play a key role also in the case $n=3$. This is the {\em
multiplicative basis of evolution}.

We denote the matrix and its elements as $\mathcal A=\pmatrix{a&b\cr c&d}$, and the
determinant of $\mathcal A$ as $\Delta=ad-bc$. The crucial point is that
both transformations $I$ and $J$ of the following six values:
\be
a, b, c, d, \Delta \hbox{ and } (-1)
\label{6values}
\ee
are described in purely multiplicative terms. Namely, $I$ act like this:
\be
a\mapsto d\Delta^{-1},\; b\mapsto (-1)b\Delta^{-1},\; 
c\mapsto (-1)c\Delta^{-1},\; d\mapsto a\Delta^{-1},\; \Delta\mapsto\Delta^{-1},\;
(-1)\mapsto (-1);
\label{I2*2}
\ee
and $J$ like this:
\be
a\mapsto a^{-1},\; b\mapsto b^{-1},\; c\mapsto c^{-1},\; d\mapsto d^{-1},\; 
\Delta\mapsto (-1)a^{-1}b^{-1}c^{-1}d^{-1} \Delta,\; (-1)\mapsto (-1).
\label{J2*2}
\ee

One can say that there are two matrices $T_I,T_J \in {\rm GL}(6,\mathbb
Z)$ which act {\em multiplicatively\/} (in the following sense: the number
2 acts on a variable~$x$ multiplicatively by $x\mapsto x^2$) on columns of values 
(\ref{6values}), where
$$
T_I=\pmatrix{ 0&0&0&1&-1&0\cr 0&1&0&0&-1&1\cr 0&0&1&0&-1&1\cr
1&0&0&0&-1&0\cr 0&0&0&0&-1&0\cr 0&0&0&0&0&1 } \quad \hbox{and} \quad
T_J=\pmatrix{ -1&0&0&0&0&0\cr 0&-1&0&0&0&0\cr 0&0&-1&0&0&0\cr 0&0&0&-1&0&0\cr
-1&-1&-1&-1&1&1\cr 0&0&0&0&0&1 }.
$$

Of course, the answer for values (\ref{6values}) after $N$ steps of evolution
is given in the similar way by the matrix~$(T_J T_I)^N$. This latter is
given by two slightly different expressions for $N$ odd and even. For example,
if $N=2k$, then
$$
(T_J T_I)^N = \pmatrix{ -k+1&-k&-k&-k&2k&-k\cr -k&-k+1&-k&-k&2k&-k\cr
-k&-k&-k+1&-k&2k&-k\cr -k&-k&-k&-k+1&2k&-k\cr -2k&-2k&-2k&-2k&4k+1&-2k\cr
0&0&0&0&0&1 }.
$$
This means that the new value of $a$ after $2k$ steps is 
\be
a(2k)=a^{-k+1}b^{-k}c^{-k}d^{-k}\Delta^{2k}(-1)^{-k},
\label{a2*2}
\ee
and similarly for $b$, $c$ and~$d$. In the r.h.s.\ of~(\ref{a2*2}), of course, the initial values
of the variables are taken.

\section{$3\times 3$ matrices: one special formula}
\label{sec 3*3 formula}

This section provides a formula necessary for building the multiplicative basis 
for the evolution of $3\times 3$ matrices, in analogy with section~\ref{sec 2*2}.

For a matrix
\be
\mathcal A=\pmatrix{ a&b&c\cr f&g&h\cr r&s&t }
\label{3*3 notation}
\ee
we denote $\dh \mathcal A$ the determinant of its Hadamard inverse multiplied,
for convenience, by the product of all elements of~$\mathcal A$:
\be
\dh \mathcal A = agtbhr + fscagt + bhrfsc - rgcfbt - fbtsha - shargc.
\label{dh}
\ee

It turns out that $\dh \mathcal A$ behaves very nicely under the (usual)
inversion of~$\mathcal A$:
\be
\dh \mathcal A^{-1} =-\dh \mathcal A\, (\det \mathcal A )^{-4}.
\label{*}
\ee

The direct proof of formula~(\ref{*}) consists simply in applying computer
algebra. This, however, does not explain how one can arrive at such formula.
In the rest of this section we present heuristic argument which clearly
shows that a formula of such kind must exist. Our argument was suggested
by paper~\cite{L}.

Let $\dh \mathcal A=0$. This means that the matrix
$$
\mathcal B =J(\mathcal A)=\pmatrix{ a^{-1}&b^{-1}&c^{-1}\cr f^{-1}&g^{-1}&h^{-1}\cr 
r^{-1}&s^{-1}&t^{-1} }
$$
is degenerate (remember that we do not care about the rigor!). This means
that $\mathcal B \pmatrix{\alpha\cr \beta\cr \gamma} =0$ for some (nonzero) column
$\pmatrix{\alpha\cr \beta\cr \gamma}$. In terms of initial matrix $\mathcal
A$, this yields:
\be
\mathcal A \pmatrix{ \alpha' &0&0\cr 0&\beta' &0\cr 0&0&\gamma' }\mathcal A^{\rm
T} =\pmatrix{ \alpha'' &0&0\cr 0&\beta'' &0\cr 0&0&\gamma'' }.
\label{**}
\ee
Here the superscript $\rm T$ means matrix transposing; the values $\alpha',
\ldots, \gamma''$ are given by
\begin{eqnarray*}
&& \alpha'=\frac{\alpha}{afr},\quad \beta'=\frac{\beta}{bgs},\quad 
\gamma'=\frac{\gamma}{cht}, \\
&& \alpha''=\alpha'a^2+\beta'b^2+\gamma'c^2,\quad 
\beta''=\alpha'f^2+\beta'g^2+\gamma'h^2,\quad \gamma''=\alpha'r^2+\beta's^2+\gamma't^2.
\end{eqnarray*}

It is clear that a relation similar to (\ref{**}) holds for $I(\mathcal
A) = \mathcal A^{-1}$ as well. This means that $(J\circ I)(\mathcal A) = 
J(\mathcal A^{-1})$ is degenerate. In other words,
$$
\dh \mathcal A =0 \; \Rightarrow \; \dh (\mathcal A)^{-1} =0.
$$

There is no need to make this argument rigorous because, as has been said,
formula~(\ref{*}) admits a direct verification.

\section{$3\times 3$ matrices: the solution}
\label{sec 3*3 solution}

We proceed along the same lines as in section~\ref{sec 2*2}, which is possible
due to formula~(\ref{*}) from  section~\ref{sec 3*3 formula}. We use
notation~(\ref{3*3 notation}) for the entries of matrix~$\mathcal A$.

The multiplicative basis of evolution comprises now 21 values: the matrix elements
$a$, $b$, $c$, $f$, $g$, $h$, $r$, $s$, $t$; the determinant
$\Delta=\det \mathcal A$; nine cofactors
of matrix elements denoted by corresponding capital letters: 
$A=\left| \matrix{g& h\cr s&t} \right|$, 
$B=-\left| \matrix{f& h\cr r&t} \right|$ and so on; the value $\Xi=\dh
A$ defined by (\ref{dh}) and, finally, the value~$(-1)$. Here is how
our two involutions act on these values.

The matrix inverse $I$:
\begin{eqnarray*}
&& a\mapsto A \Delta^{-1}, \; b\mapsto F \Delta^{-1}, \; 
\ldots, \; h\mapsto H \Delta^{-1}; \\
&& A\mapsto a \Delta^{-1}, \; B\mapsto f \Delta^{-1}, \; 
\ldots, \; H\mapsto h \Delta^{-1}; \\
&& \Delta \mapsto \Delta^{-1}, \quad \Xi \mapsto (-1) \Xi \Delta^{-4}, \quad  
(-1)\mapsto (-1).
\end{eqnarray*}

The element-by-element inverse $J$:
\begin{eqnarray*}
&& a\mapsto a^{-1}, \; \ldots, \; h\mapsto h^{-1}; \\
&& A\mapsto (-1)A g^{-1}h^{-1}s^{-1}t^{-1}, \; \ldots, \; H\mapsto (-1)H 
a^{-1}b^{-1}f^{-1}g^{-1}; \\
&& \Delta \mapsto \Xi a^{-1}b^{-1}c^{-1}f^{-1}g^{-1}h^{-1}r^{-1}s^{-1}t^{-1}, \\
&& \Xi \mapsto \Delta a^{-1}b^{-1}c^{-1}f^{-1}g^{-1}h^{-1}r^{-1}s^{-1}t^{-1}, \\
&& (-1)\mapsto (-1).
\end{eqnarray*}

A matrix $T_J T_I$ corresponding to a step of evolution can now be calculated
in analogy with section~\ref{sec 2*2}, but now it has sizes $21\times 21$,
and we do not write it out here. Still, it makes no difficulty for a computer
to handle such matrices. The remarkable fact is that {\em all the eigenvalues
of\/ $T_J T_I$ are sixth roots of unity}, as the following table shows:
$$ 
\begin{array}{|c|c|c|c|c|c|c|}
\hline 
\hbox{eigenvalues}\vphantom{\Big|}
&\frac{1+\sqrt{-3}}{2}&\frac{1-\sqrt{-3}}{2}&
\frac{-1+\sqrt{-3}}{2}&\frac{-1-\sqrt{-3}}{2}&1&-1\\ \hline
\hbox{multiplicity}&1&1&4&4&7&4\\ \hline
\end{array}
$$
Of course, $T_J T_I$ does have nontrivial Jordan boxes. An interesting
thing with them is that they all correspond only to eigenvalues~$\pm 1$. 

Here is the explicit answer for the variable~$a$ after $3k$~steps:
\be
a(3k)=a^{-2k+1} b^{-k} c^{-k} f^{-k} r^{-k} \Delta^{2k}
A^{-2k} B^{-k} C^{-k} F^{-k} R^{-k} \Xi^{2k} (-1)^{-3k(2k-1)}.
\label{otvet a 3*3}
\ee
Again, in the r.h.s.\ of formula~(\ref{otvet a 3*3}) the {\em initial\/} values of
all variables are taken.

\section{$4\times 4$ matrices: the conservation laws and computer algebra results}
\label{sec 4*4 computer}

In this section we describe the way that has actually led to the ansatz
presented in subsequent sections. Formally, however, the construction of
our ansatz does not rely on the material of this section.

First, the evolution of a $4\times 4$ matrix $\mathcal A$ can be considered
to within the following {\em gauge freedom}: we can consider not $\mathcal A$
itself but its equivalence class with respect to its multiplication both
from the left and from the right by some diagonal matrices $\mathcal B$
and~$\mathcal C$. Clearly, if 
\be
\mathcal A'=\mathcal B\mathcal A\mathcal C
\label{BAC}
\ee
then
\be
(J\circ I)(\mathcal A')=\mathcal C (J\circ I)(\mathcal A)\mathcal B.
\label{10a}
\ee
Using a transformation~(\ref{BAC}), we can reduce (almost any) $\mathcal A$ 
to the following form:
\be
\mathcal A = \pmatrix{1&1&1&1\cr 1&a&b&c\cr 1&f&g&h\cr 1&r&s&t}.
\label{1*}
\ee
So, in a sense, there remains in $\mathcal A$ nine parameters. If we find
eight conservation laws this will be a strong argument suggesting that
the evolution goes just along elliptic curves (because only elliptic and
rational curves have infinite number of automorphisms).

Second, there {\em are\/} many conserved quantities which are, moreover,
invariant under~(\ref{BAC}). Consider a decomposition of a $4\times 4$
matrix in four $2\times 2$ submatrices, for instance, one of the following:
$$
\pmatrix{ \diamond &\diamond&*&*\cr \diamond &\diamond&*&*\cr 
\| &\| &\S &\S\cr \| &\| &\S &\S} \quad \hbox{or} \quad
\pmatrix{ \diamond &\S &\diamond &\S\cr \| &* &* &\| \cr
\diamond &\S &\diamond &\S\cr \| &* &* &\| },
$$
where the entries denoted by the same symbol belong to the same submatrix.

For any such decomposition $p$, we construct the value $\Pi_p$ --- the
product of four corresponding minors. One can verify that under one step
of evolution, the ratio of any two such values $\Pi_p / \Pi_q$ goes into
$\Pi'_{p} / \Pi'_{q}$, where $\Pi'_{p}$ is the product of the {\em cofactors\/}
of the minors of decomposition~$p$ (in the {\em new\/} matrix~$\mathcal
A$). So, $\Pi_p / \Pi_q$ is invariant under {\em two\/} steps
of evolution.

Computer experiments show that there are eight algebraically independent
invariants of such kind. Moreover, given fixed values of these invariants and using
the form~(\ref{1*}) for the matrix, one can exclude all except two parameters
in~(\ref{1*}) from the equations and get the curve as given by just one
equation in two variables. Its genus turns out to be one, as expected. Besides,
one can see from that equation that, for example, there are four
points in the curve where the function~$a(z)$ ($z$ being a parameter for
the elliptic curve) takes value~$1$, and the function~$b(z)$ takes value~$1$
in exactly the same points, as well as some information about the coincidence
of some poles and zeros of those functions.

This was exactly what led us to the ansatz presented in the following
sections.

\section{A determinant of theta-function ratios}
\label{sec theta}

The key formula for our ansatz is the formula for the determinant of the
$n\times n$ matrix
\be
\mathcal K = (k_{ij}), \quad \hbox{where} \quad k_{ij} = 
\frac{\vartheta (y-\lambda_i-\mu_j)}{\vartheta (x+\lambda_i+\mu_j)}.
\label{calK}
\ee
Here $i$ stays of course for the number of a row and $j$ for the number of a column.
So, there are complex variables $x$ and~$y$ and, moreover, two arrays, 
$(\lambda_i)$ and $(\mu_j)$, each of $n$ complex variables. By $\vartheta$
we denote here the odd Jacobian theta-function:
$$
\vartheta(u)=2q^{1/4} \sin \frac{\pi u}{2K} \prod_{n=1}^{\infty}
(1-2q^{2n}\cos \frac{\pi u}{K}+q^{4n})(1-q^{2n}),
$$
where $q=\exp(-\pi K'/K)$; $K$ and $K'$ are called {\em half-periods}.
Our formula states that\vadjust{\goodbreak}
\begin{eqnarray}
\det \mathcal K = - \Bigl(\vartheta(x+y)\Bigr)^{n-1} \vartheta\Bigl((n-1)x-y+
\sum_{i=1}^n \lambda_i +\sum_{j=1}^n \mu_j\Bigr)\nonumber \\ 
\cdot \, \frac{\displaystyle \prod_{1\le i_1<i_2 \le n} 
\vartheta(\lambda_{i_2}-\lambda_{i_1}) \cdot \prod_{1\le j_1<j_2 \le n} 
\vartheta(\mu_{j_2}-\mu_{j_1})} {\displaystyle \prod_{i=1}^n  \prod_{j=1}^n  
\vartheta(x+\lambda_i+\mu_j)}\,.
\label{vf}
\end{eqnarray}

{\em Proof\/} of formula~(\ref{vf}) goes by induction on~$n$. Consider
$\det \mathcal K$ as a function of~$\lambda_n$. It is obliged to have the
following form:
\be
\det \mathcal K = F\, \frac{\vartheta(\lambda_n-G)\prod_{i=1}^{n-1}
\vartheta(\lambda_n-\lambda_i)}{\prod_{j=1}^{n}\vartheta(x+\lambda_n+\mu_j)},
\label{p1}
\ee
where $F$ and $G$ are some quantities that do not depend on $\lambda_n$
but can depend on other variables. Here are the reasons why exactly such
factors appear in formula~(\ref{p1}), typical for proofs of the like formulas for 
theta-functions. 

Each factor $\vartheta(\lambda_n-\lambda_i)$ in the numerator is
responsible for the fact that $\det \mathcal K$ obviously vanishes when
$\lambda_n$ coincides with any other~$\lambda_i$ (because of two
identical rows). The denominator of~(\ref{p1}) is simply the common
denominator of all elements of~$\mathcal K$. Finally, the
factor~$\vartheta(\lambda_n-G)$ is necessary to ensure that the whole
expression has the same number of zeros and poles as a function of~$\lambda_n$.

On the other hand, in the neighborhood of the value $\lambda_n=-x-\mu_n$,
where $k_{nn}$ has a pole, our determinant behaves as
\be
\det \mathcal K \approx \det \mathcal K_{\rm smaller} 
\frac{\vartheta(x+y)}{\vartheta(x+\lambda_n+\mu_n)},
\label{p2}
\ee
where $\mathcal K_{\rm smaller}$ is the same matrix $\mathcal K$ but {\em
without\/} its $n$th row and $n$th column. Comparing (\ref{p2}) with (\ref{p1})
(where we at this moment also substitute $\lambda_n=-x-\mu_n$), we can get the 
quantity~$F$ in the following form:
\be
F= \det \mathcal K_{\rm smaller} \frac{\vartheta(x+y) \prod_{j=1}^{n-1}
\vartheta(\mu_n+\mu_j)}{\vartheta(-x-\mu_n-G) \prod_{i=1}^{n-1}
\vartheta(-x-\mu_n-\lambda_i)}.
\label{p3}
\ee
Now, substituting (\ref{p3}) in (\ref{p1}), we get the following, almost
final, formula expressing $\det \mathcal K$ through $\det \mathcal K_{\rm smaller}$:
\begin{eqnarray}
&& \det\mathcal K = \det\mathcal K_{\rm smaller} \nonumber \\[\smallskipamount]
&& \cdot \frac{\vartheta(x+y) \prod_{j=1}^{n-1}
\vartheta(-\mu_n+\mu_j) }{\vartheta(-x-\mu_n-G) \prod_{i=1}^{n-1}
\vartheta(-x-\mu_n-\lambda_i)} \cdot \frac{\vartheta(\lambda_n-G) \prod_{i=1}^{n-1}
\vartheta(\lambda_n-\lambda_i)}{ \prod_{j=1}^{n} \vartheta(x+\lambda_n+\mu_j)}.
\label{p4}
\end{eqnarray}

It remains to calculate the value~$G$. It can be deduced from the following
reasoning. According to the inductive hypothesis, $\det\mathcal K_{\rm smaller}$
contains the multiplier $\vartheta\Bigl((n-2)x-y+\sum_{i=1}^{n-1}\lambda_i+
\sum_{j=1}^{n-1}\mu_j\Bigr)$, so it must have a zero at such $\lambda_n$ when
the argument of that theta-function equals zero. On the other hand, $\det\mathcal K$,
generally, does {\em not\/} have a zero at such~$\lambda_n$. So, the mentioned
theta-function must cancel with the same factor in the denominator of~(\ref{p4}),
which role only $\vartheta(-x-\mu_n-G)$ can assume. This implies
$$
-x-\mu_n-G = \pm \Bigl((n-2)x-y+\sum_{i=1}^{n-1}\lambda_i+
\sum_{j=1}^{n-1}\mu_j\Bigr),
$$
and the sign here, namely plus, can be fixed for example by considering
the expression~(\ref{p4}) as a function of~$x$ (which must have the right
difference, namely $n$, between the number of poles and zeros in a parallelogram
of periods).

Once we have got the proper formula for transition from $\det\mathcal K$ to
$\det\mathcal K_{\rm smaller}$, the inductive step is over. As for the
induction basis, the formula~(\ref{vf}) does obviously hold for $n=1$.

\section{Expression for a matrix element of $(J\circ I)\mathcal A$}
\label{sec matrix element}

Let us apply the transformation $J\circ I$, i.e.\ one step of our evolution,
to the $n\times n$ matrix $\mathcal A = \mathcal K$ given by
formula~(\ref{calK}). A matrix element $a_{ji}^{\rm new}$
of the obtained matrix $\mathcal A^{\rm new}$ is given by formula
\be
a_{ji}^{\rm new}=\frac{\det\mathcal A}{A_{ij}},
\label{aji}
\ee
where $A_{ij}$ is the cofactor for the element $a_{ij}$ of~$\mathcal A$.
Both the numerator and denominator in~(\ref{aji}) are determinants of the
form~(\ref{vf}). The calculation yields:
\be
a_{ji}^{\rm new}=a^{\rm global}a_{j}^{\rm row}
a_{i}^{\rm column}a_{ij}^{\rm element},
\ee
where
\be
a^{\rm global} = \vartheta(x+y)\vartheta\Bigl((n-1)x-y+
\sum_{k=1}^n\lambda_k + \sum_{l=1}^n \mu_l\Bigr)
\label{a global}
\ee
is a factor which depends neither on $j$ nor on $i$;
\be
a_{j}^{\rm row}=\frac{\prod_{l=1,\,l\ne j}^n \vartheta(\mu_j-\mu_l)}{
\prod_{k=1}^n \vartheta(x+\lambda_k+\mu_j)}
\label{aj row}
\ee
depends only on $j$ (the row number for $a_{ji}^{\rm new}$);
\be
a_{i}^{\rm column}=\frac{\prod_{k=1,\,k\ne i}^{n-1}
\vartheta(\lambda_i-\lambda_k)}{\prod_{l=1}^n \vartheta(x+\lambda_i+\mu_l)}
\label{ai column}
\ee
depends only on $i$ (the column number); and the last factor
\be
a_{ji}^{\rm element}= \frac{\vartheta(x+\lambda_i+\mu_j)}{\vartheta
\Bigl((n-2)x-y+\sum_{k=1}^n \lambda_k +\sum_{l=1}^n \mu_l -
\lambda_i-\mu_j \Bigr)}
\label{aji element}
\ee
has much the same form as the initial $a_{ij}$.

To be exact, the difference between the matrix $(a_{ji}^{\rm element})$
made of matrix elements (\ref{aji element}) and the initial
matrix~$\mathcal A=\mathcal K$ can be described as follows: change
$$
x\mapsto x^{\rm new}=y-(n-2)x-\sum_{i=1}^n \lambda_i -\sum_{j=1}^n \mu_j,
\quad y\mapsto y^{\rm new}=-x,
$$
and then perform the matrix transposing. As for the factors (\ref{aj row})
and (\ref{ai column}), their effect consists in multiplying
the matrix $(a_{ji}^{\rm element})$ from two sides by diagonal matrices,
i.e.\ doing a gauge transformation~(\ref{BAC}). The main point is that
if we have done not one but $N$ steps of evolution, the effect of {\em
all\/} arising factors (\ref{aj row}) and (\ref{ai column}), as well as
(\ref{a global}), consists just in the appearing of some products of
theta-functions with their arguments changing according to a simple law.
We do not write out here the corresponding obvious but bulky formulas.
What we see already is that {\em the evolution of a matrix of the
form~(\ref{calK}) can be described by an explicit formula}. The same
applies, obviously, to a matrix that was obtained from such one by
a transformation~(\ref{BAC}).

{\bf Remark. \vadjust{\nobreak}}Here we do {\em not\/} mean just the
evolution of gauge equivalence classes as in section~\ref{sec 4*4 computer}.
Formula~(\ref{10a}) shows that if we know what happens with $\mathcal A$,
we also know what happens with $\mathcal B\mathcal A \mathcal C$ after any
number of evolution steps.

\section{Comparing different values of $n$}
\label{sec disc}

\subsection{$n=3$}

The ansatz (\ref{calK}) (taken together with the possibility of multiplying
a matrix by two diagonal matrices as in formula~(\ref{BAC})) is definitely
superfluous for the case $n=3$ where we have presented, in sections
\ref{sec 3*3 formula} and~\ref{sec 3*3 solution}, a more direct approach
which, by the way, gives the exhaustive information about the cases where
the evolution cannot go ahead because of a division by zero.

\subsection{$n=4$}
\label{subsec n=4}

Here the ansatz (\ref{calK}) together with multiplication by two diagonal matrices contains
exactly 16 independent parameters, i.e.\ gives a Zariski open set of
matrices. To see this, note first that the 10 values $x$, $y$, $\lambda_i$
and $\mu_j$ produce really only 8 parameters, because nothing in (\ref{calK})
changes if we do one of the following translations ($\alpha$ being an
arbitrary complex number):
$$
\lambda_i \mapsto \lambda_i+\alpha \hbox{ for all }i, \quad
\mu_j \mapsto \mu_j - \alpha \hbox{ for all }j
$$
or
$$
x\mapsto x+\alpha,\quad y\mapsto y-\alpha, \quad
\lambda_i \mapsto \lambda_i-\alpha \hbox{ for all }i.
$$
Second, the {\em modulus of the elliptic curve\/} is the 9th parameter.
And finally, the two diagonal matrices produce the 7 remaining parameters.

\subsection{$n\ge5$}

Our ansatz works for any~$n$, but when $n>4$ it corresponds only
to a subvariety of a nonzero codimension in the space of all
$n\times n$ matrices. For instance, for $n=5$, the calculation of parameters
similar to that done in subsection~\ref{subsec n=4} shows that we can
describe in the same way the evolution of a 20-parameter family of matrices.

\section{A special case with period 4}
\label{sec period 4}

Although the parameterisation~(\ref{calK}) together with the possibility
of multiplying the matrices by two diagonal ones as in~(\ref{BAC})
encompasses in the case $n=4$ a Zariski open subset of all $4\times 4$
matrices, it does not include some interesting special cases (and it
does not seem very easy to obtain them as any limiting cases). One
specific feature of parameterisation~(\ref{calK}) is that {\em if some
$2\times 2$ minor of the matrix equals zero, then all other minors also do
so}. We present here a matrix 
\be
\mathcal A=\pmatrix{ 1&1&1&1\cr 1&1&xy&x\cr 1&zt&1&z\cr 1&t&y&1 }
\label{period 4}
\ee
that does not obey such a requirement. Its interesting property is that
if we consider its evolution {\em up to gauge transformations~(\ref{BAC})},
then after four steps we get back at the initial matrix~(\ref{period 4}).
This statement is proved by a direct computer calculation.

\medskip

{\bf Acknowledgements. }I thank Claude Viallet for informing me about the
paper~\cite{AMV}. This work has been performed with the partial financial
support from Russian Foundation for Basic Research, Grant no.~01-01-00059.


\begin{thebibliography}{99}

\bibitem{diss}
I.G. Korepanov.
Algebraic integrable dynamical systems, 2+1-dimensional models in wholly discrete 
space-time, and inhomogeneous models in 2-dimensional statistical physics.
ArXiv: solv-int/9506003.

\bibitem{AMV} 
J.-Ch. Angl\`es d'Auriac, J.-M. Maillard and C.M.~Viallet.
A classification of four-state spin edge Potts models.
ArXiv: cond-mat/0209557.

\bibitem{L} 
J.M. Landsberg.
On an unusual conjecture of Kontsevich and variants of Castelnuovo's lemma.
ArXiv: alg-geom/9604023.

\end{thebibliography}
\end{document}